\documentstyle[eqsecnum,aps,epsf,twocolumn]{revtex}

\begin{document}

\preprint{DUKE-TH-99-184}

\title{
       Real-Time Evolution of Soft Gluon Field Dynamics in 
       Ultra-Relativistic Heavy-Ion Collisions
      }

\author{W. P\"oschl and B. M\"uller   \\
Department of Physics, Duke University, Durham, NC 27708-0305, USA }
\vspace{15mm}
\date{\today}
\maketitle
\vspace{15mm}

\maketitle

\begin{abstract}
The dynamics of gluons and quarks in a relativistic nuclear      
collision are described,
within the framework of a classical mean-field transport theory,
by the coupled equations for the Yang-Mills field and a collection
of colored point particles. The particles are used to represent 
color source effects of the valence quarks in the colliding nuclei. 
The possibilities of this approach are studied
to describe the real time evolution of small $x$ modes in the
classical effective theory in a non-perturbative coherent manner. 
The time evolution of the color fields is explored
in a numerical simulation of the collision of two 
Lorentz-boosted clouds of color charged particles 
on a long three-dimensional gauge lattice. 
We report results on soft gluon scattering and coherent gluon radiation
obtained in SU(2) gauge symmetry.
\end{abstract}
\bigskip\bigskip
%
%
%
%
%
\section {Introduction}
\noindent 
Experiments with relativistic heavy ions at center-of-mass energies reaching
100 GeV/u will soon search for a new phase of nuclear matter
\cite{Harris.96,Smilga.97}. 
It was argued that the extraordinarily high energy and particle number
densities reached in central nuclear collisions at RHIC \cite{Gyulassy.95}
could lead to rapid local thermalization of matter \cite{Shuryak.92}
and thus to the formation of the so-called quark gluon plasma 
\cite{Muller.95}. 

\noindent 
One of the theoretical challenges in this 
context is the development of a description on the basis of quantum 
chromodynamics (QCD) of the processes that may lead to the formation 
of the locally equilibrated
quark gluon plasma state in these nuclear reactions.
Predictions on whether and how this state could be formed
in a nuclear collision depend in a sensitive way on the initial conditions.
At present, our understanding of these conditions is rather poor. 
Perturbative parton-parton scattering 
precedes local thermo-dynamical equilibrium in the early stage 
and may partly determine the dynamics of thermalization. Indeed, 
event generators based on individual parton-parton processes
\cite{Geiger.92} have been developed. However, their 
predictions differ widely. To improve our understanding of the
early stage, we have to consider the fact that the initial state
is described by coherent parton wave functions and glue fields.
Consequently, coherent multiple scattering must play a role in
the early stage of a collision. 
The simplest approach to this picture would be to start with
a mean field description of the color fields in an effective theory
while hard modes are described through classical particles representing
partons. 

\noindent 
The possibility of a description of inelastic gluon processes by means
of the nonlinear interactions of classical color fields has been proposed
some time ago \cite{Ehtamo.83,Kovner.95}. Some years ago, this scenario was
examined in studies of the collision of two transverse polarized Yang-Mills 
field wave packets on a one-dimensional gauge lattice \cite{Hu.95}. These
calculations showed that the interaction between localized classical 
gauge fields can lead to the excitation of long wavelength modes in a way 
that is reminiscent of the formation of a dense gluon plasma. Recently,
a similar study has been carried out for collisions of Yang-Mills wave
packets on a three-dimensional gauge lattice \cite{Po.98}.

\noindent 
Here, we go a step further and add color charged particles to the system
acting as sources of the color fields. Instead of two wave packets, we
collide two clouds of such particles which shall describe the
color charge distributions generated by the valence quarks in
two colliding nuclei.
The wave packets are now replaced by the colliding gluon fields 
which are generated by classical color sources
while they propagate through the lattice, 
representing the fast moving nuclear valence quarks. 
The inclusion of 
particles also permits a comparison with perturbative calculations of 
gluon radiation from colliding quarks \cite{Kovchegov.97,Matinyan.97}.

\noindent 
We address the question whether indications for the formation of
a quark gluon plasma can also be observed in such
simulations.
For simplicity, we leave out
the hard collisions between the particles and focus first only 
- except from the time dependence of the color sources - on
the soft dynamics in the gluon fields which are
described on the lattice and correspond to modes with a small
light cone momentum fraction $x$. We further
leave out the hard modes of the gluon fields which may also
be represented by particles. The particles
may in principle be used in the future to include the semi-hard dynamics 
at larger $x$ following the idea of the parton cascade model 
\cite{Geiger.95}.
While this ``probabilistic'' approach provides a reasonable
description of large transverse momentum processes at large $x$,
QCD coherence effects become important as we go to small $x$ or
alternatively, towards central rapidities \cite{Venu.98}.     
What is needed therefore to describe the collision of the
``wee'' partons is a wave picture where coherent multiple
scattering is fully taken into account.

\noindent 
At the limit of extremely high collision energies, the model
presented here has in principle to converge against
a random light-cone source model which was
initially proposed by McLerran and Venugopalan \cite{Venu.94} and
later further developed by these authors in collaboration with
Ayala, Jalilian-Marian, Kovner,
Leonidov, and Weigert
\cite{Ayala.95,Jalilian.1,Jalilian.2,Jalilian.3}. 
This model may therefore be used as a reference in the 
limit ${v/c} \rightarrow 1$. The McLerran-Venugopalan model
considers very large nuclei moving with nearly the speed of light which
consequently appear in the laboratory frame as ``pancakes'' of almost zero
thickness in the transverse plane. The classical gluon field at
small $x$ for a nucleus in the infinite momentum frame is obtained
by solving the Yang-Mills equations in the presence of a static
source of color charge on the light cone. 
In our approach, the solution of the equations of motion requires
one to take the longitudinal extension of the nucleus into account,
i.e. one must not take the nucleus to be infinitely thin in the
longitudinal direction as assumed in \cite{Venu.94}.
Following Kovchegov and Rischke \cite{Rischke.97}, we therefore 
take the finite longitudinal extension of the nuclei into account.
This of course requires that the velocities of the nuclei are
slightly less than the speed of light.

\noindent 
A major goal of the description, presented in this paper, 
is to keep the possibility open for an inclusion of parton cascades
in the further development while the coherent physics  
of the random light-cone source model at small $x$ 
is described at the same time.
This goal can only be accomplished in several steps.
The approach presented here contains no restriction to
central rapidities and provides therefore an extension to
larger x taking into account semi-hard gluons.

\noindent 
In section 2, we present a brief description of the model
and the formulation on a SU(2) gauge lattice.
In section 3, we discuss the preparation of the initial state
of the nuclei and show numerical results. In section 4, we 
evolve the collision and present results for the time evolution
of the glue fields in the collision. A summary and conclusion
is given in the section 5.
%
%
%
%
%
\section {Description of the Model}
\noindent 
A set of equations describing the evolution of the phase space
distribution of quarks and gluons in the presence of a mean color field,
but in the absence of collisions, was proposed by 
Heinz \cite{Heinz.83,Elze.89}. 
This non-Abelian generalization of the 
Vlasov equation can be considered as the continuum version of the dynamics
of an ensemble of classical point particles endowed with color charge
and interacting with a mean color field. Denoting the space-time positions,
momenta, and color charges of the particles by $x_i^{\mu}$, $p_i^{\mu}$ and 
$q_i^a$, respectively, where $i=1,\ldots,N$ is the particle index, the
equations \cite{Wong.70} for this dynamical system read:
\begin{eqnarray}
\label{Eq.1}
m{{dx_i^{\mu}}\over{d\tau}} &=& p_i^{\mu} \\
\label{Eq.2}
m{{dp_i^{\mu}}\over{d\tau}} &=& gq_i^aF^{\mu\nu}_a p_{i,\nu} \\
\label{Eq.3}
m{{dq_i^{a}}\over{d\tau}}   &=& - g f^{abc} q_i^b p_i^{\mu} A_{\mu}^c .
\end{eqnarray}
The moving particles generate a color current ${\cal J^{\mu}}=J^{\mu}_a
\tau^a/2$ which forms the source term of the inhomogeneous Yang-Mills 
equations 
\begin{eqnarray}
\label{Eq.5}
\bigl[{\cal D}_{\mu},{\cal F}^{\mu\nu}(x)\bigr]& = &
 g\,{\cal J}^{\nu}(x) \nonumber\\
&=&
g\,\sum_i {\cal Q}_i(t) {{p_i^\nu}\over{m}} \delta(\vec x-\vec x_i(t)) ,
\end{eqnarray}
for the mean color field. 
These equations have recently been used in the weak-coupling limit 
($g\ll 1$) to simulate the effects of hard
thermal loops \cite{Pisarski.89} on the dynamics of soft modes of a
non-Abelian gauge field at finite temperature \cite{Hu.97,Moore.98}.

\noindent 
Here we use Eq. (\ref{Eq.1}--\ref{Eq.5}) to describe the 
interactions among the soft glue field components of two colliding heavy 
nuclei. Transverse modes are included by using a 3-dimensional spatial
gauge lattice. The short-distance lattice cut-off $a$ separates the
regime in transverse momentum where the dynamics of gluons is 
perturbative
(large $k_{\rm T}$) from that where perturbation theory fails (small 
$k_{\rm T}$) and which is of interest to us here.
The interaction with the mean color field allows for an exchange of an 
arbitrary number of soft gluons. In combination with a parton cascade,
the screening of the soft components of the gauge field by 
perturbative partons \cite{Biro.92,Eskola.96} is taken into 
account naturally by the nonlinear
nature of the coupled Eq. (\ref{Eq.1}--\ref{Eq.5}). This will be
discussed in a separate publication \cite{BMP.98}.

\noindent 
We represent the valence quarks of the 
two colliding nuclei as point particles moving in the space-time continuum,
and interacting with a classical gauge field defined on a spatial lattice
in continuous time. Here we neglect the collision integrals describing 
hard interactions between the particles. In the spirit of the statistical
nature of the transport theory, we split each quark into a number $n_q$ of 
test particles, each of which carries the fraction $q_0=Q_0/n_q$ of the 
quark color charge $Q_0=\sqrt{3/4}$. For the gauge group SU(2) adopted here, 
each nucleon is represented by two quarks, initially carrying 
opposite color charge.  A lattice version of the continuum equations 
(\ref{Eq.1}--\ref{Eq.5}) is constructed 
\cite{Hu.97} by expressing the field amplitudes as elements of the 
corresponding Lie algebra, i.e. ${\cal A}_{\mu},{\cal F}_{\mu\nu},
{\cal E}_k,{\cal B}_k\,\in$ LSU(2).  As in the Kogut-Susskind model 
of lattice gauge theory \cite{Kogut.75} we choose the temporal gauge 
${\cal A}_0 = 0$ and define the following variables. 
\begin{eqnarray}
\label{Eq.6}
{\cal U}_{x,l} 
&=& \exp(-iga_l{\cal A}_l(x)) \,\,=\,\, {\cal U}^{\dagger}_{x+l,-l} \\
\label{Eq.7}
{\cal U}_{x,kl} 
&=& {\cal U}_{x,k}\, {\cal U}_{x+k,l}\, {\cal U}_{x+k+l,-k}\, 
                  {\cal U}_{x+l,-l} 
\end{eqnarray}
Consequently, we have
\begin{eqnarray}
\label{Eq.8}
{\cal E}_{x,j} &=& { {-i}\over{ga_j}}\,\dot {\cal U}_{x,j}
                   {\cal U}^{\dagger}_{x,j}, \nonumber\\
{\cal B}_{x,j} &=& { {i\,\epsilon_{jkl}}\over{4ga_ka_l}}\,
\bigl({\cal U}_{x,kl} - 
{\cal U}^{\dagger}_{x,kl} \bigr),
\end{eqnarray}
for the electric and magnetic fields, respectively. The lattice constant
in the spatial direction $j$ is denoted by $a_j$.  As one can see from 
(\ref{Eq.6}), the gauge field is expressed in terms of the link variables 
${\cal U}_{x,l}\,\epsilon$ SU(2), which represent the parallel transport 
of a field amplitude from a site $x$ to a neighboring site $(x+l)$ in 
the direction $l$.  We choose ${\cal U}_{x,k}$ and ${\cal E}_{x,k}$ as the 
basic dynamic field variables and solve the following equations of motion
for the fields.
\begin{eqnarray}
\label{Eq.10}
\dot{\cal U}_{x,k}(t)& =& i\,g\,a_k\,{\cal E}_{x,k}(t)\,{\cal U}_{x,k}(t) 
\qquad\qquad\qquad\qquad\qquad \\
\label{Eq.11}
\dot{\cal E}_{x,k}(t) &=& 
{i\over{2ga_1a_2a_3}}
\sum\limits^3_{l=1}
\Bigl\{{\cal U}_{x,kl}(t) -
{\cal U}^{\dagger}_{x,kl}(t) 
\,\Bigr.  \nonumber\\
&-&\,  \Bigl. {\cal U}^{\dagger}_{x-l,l}(t)\,
\Bigl( {\cal U}_{x-l,kl}(t) -
{\cal U}^{\dagger}_{x-l,kl}(t)\Bigr)\,
{\cal U}_{x-l,l}(t)\, \Bigr\} \nonumber\\
&-&\,g\,{\cal J}_{x,k}(t).
\end{eqnarray}
In the time evolution of the particle lattice system,
these equations have to be solved simultaneously
together with the equations (\ref{Eq.1}) to (\ref{Eq.3}).
The dynamics of the particles is described within a
dual lattice which is defined by the edges of 
the Wigner-Seitz cells of the original lattice. 
The charge of all particles in the volume of one
cell is associated with the center of the cell represented
by a point of the real lattice. As long as a particle propagates
in the cell its charge vector is not changed. If a particle 
crosses the boundary of a dual box, the charge is transported
parallel to the center of the neighbor cell into which
the particle enters. Exchange of energy between a particle
and the field occurs only at the transition from one dual
box into another. The particle is reflected from the wall
if its kinetic energy is smaller or equal than the work required
to move the charge from the old box into the new box.
In the other case, a transition into the new box occurs
where the amount of energy required to make the transition
is either subtracted or added to the kinetic energy of the
particle depending on the sign of the product 
$Tr({\cal Q}_i{\cal E}_{x,l})$.
This mechanism provides a coupling between the fields and
the particles which allows to generate color fields
through color charge currents. The method has the advantage
that it exactly preserves the law of Gauss. It also preserves
the conservation of energy which is thus only violated through
numerical errors of the order $\Delta t^3$ in the time integration
of the field equations (\ref{Eq.10}) and (\ref{Eq.11}).
On the other hand, the method, as simple as it is, is connected with
serious limitations as for example the violation of causality
when the time step width $\Delta t$ is smaller than the lattice
constants $a_l$. $\Delta t$ can not be chosen larger than $a_l$ 
because the time integration of the equations (\ref{Eq.10}) and 
(\ref{Eq.11}) is instable for $\Delta t \ge {\rm min}(a_l:l=1,2,3)$
These limitations do not yet allow a study of the dynamics of the
particles in a microscopically correct way. Artifacts, as
for example a trapping of particles of too small kinetic energy
in dual cells can occur. 
For propagating nuclei, the field energy associated with
a single link can occasionally exceed the kinetic energy of 
a particle resulting in the emission of particles due to backwards
scattering from the link. This corresponds to the (unphysical) 
emission of nucleons from the moving nuclei, which we want to avoid.

\noindent 
However, a slight modification of the above described procedure
for the wall transition of particles allows to use the particles
to simulate color charge fluctuations and to generate coherent
color fields: We leave
the kinetic energy of a particle unchanged whenever it crosses
a cell boundary of the dual lattice and only transport its color charge
to the neighboring lattice cell. This is equivalent to setting
the Coulomb force in the r.h.s. term of Eq. (\ref{Eq.3}) to zero. 
In this way, the collection
of particles acts as a classical color source 
propagating through the lattice
with constant velocity.
%
%
%
%
%
\section {The Initial State}
\noindent 
In the following, we briefly outline how, in principle,
a simulation of a nuclear collision is performed within the model 
described in the previous section. Mainly, however, we discuss the
question how to generate the ``initial state''.

\noindent 
For a central collision of two Pb nuclei one would expect
that 34 nucleons in a row collide on the collision axis
(z-axis). 
This number is desired from the following estimate.
We denote the nucleon radius by $r_n$, the proton radius
of a nucleus by $R_P$ and the neutron radius of a nucleus by $R_N$.
The average number of nucleons in a cylinder of the radius
$r_n$ around the collision axis is determined by
\begin{equation}
\label{Eq.3.1}
n=n_P+n_N={3\over 2}\Bigl( \Bigl({{r_n}\over{R_P}}\Bigr)^2 Z +
                           \Bigl({{r_n}\over{R_N}}\Bigr)^2 N \Bigr).
\end{equation}
With the values $r_n=1.32\,{\rm fm}$, $R_P=5.47\,{\rm fm}$,
$R_N=5.80\,{\rm fm}$ from Ref. \cite{Ring.89}, we 
find the result $n=17.11$ for $^{208}Pb$.

\noindent 
In accordance with the spatial extension of a nucleon
we choose a lattice extension of 1.2~fm into both transverse 
directions. The lattice spacing is taken $a_l = 0.3$ fm in
each direction, thus $4^2\times 40$ lattice points cover the volume 
of 17 nucleons in one row. More correctly, since there is no 
Pauli-Principle between
neutrons and protons there are two superposed rows of nucleons.
One row contains 8 protons and the other 9 neutrons. These
details, however, are far beyond the reach of our calculation,
and we simply take 16 nucleons or 32 quarks (in SU(2))
in our initial state. We also use the rounded value
$r_n=1.2$ fm.
The coverage of two complete nuclei in the transverse plane
requires much larger lattices and remains a challenge for more
extended calculations in the future. 
The lattice is closed to a 3-torus which means that we apply
periodic boundary conditions in all three spacial directions.
The fact that we use a lattice with transverse extensions
much smaller than the radius of a nucleus leads to restrictions.
The first of these is that we can not simulate the full transverse
dynamics of a collision. Such simulations for central collisions
are possible but they require lattices
with transverse extensions at least twice as large as the radius
of a nucleus. Second, for lattices with small transverse extensions,
the periodic boundary conditions in the transverse directions 
define a situation which is equivalent to the assumption that
the nuclei have infinite extension into transverse directions. 
Therefore, in this respect, our description stays firmly within 
the framework of the model of McLerran and Venugopalan \cite{Venu.97}.
Larger lattice sizes will be necessary to carry
out a meaningful Fourier analysis of the fields in transverse directions
and to compare results of both models. 
However, the transverse extension
of a nucleon can be covered at least to describe the initial
state of nucleons.

\noindent 
As already mentioned above, 
a dual lattice is superimposed on the original lattice in such 
a way that the lattice points are located in the centers of the cells of 
the dual lattice. 
The dual lattice cells are used to associate particles
with lattice sites and thus to define the source terms in (\ref{Eq.5}).
To generate the initial configurations of the nuclei, we randomly distribute 
color charged massless particles over the volume such that each lattice cell
is occupied by an even number $n_b$ of particles. The total initial color
charge is zero in each box corresponding to a neutral color 
charge distribution.
Momenta with opposite but random directions and Boltzmann distributed 
absolute values are assigned to each pair of particles
\footnote{
In the ``real'' initial state of a nucleus, each nucleon 
of any shell is in its
ground state with a temperature $T=0$. On the other hand each
nucleon has a finite energy according to its rest mass. Only a 
full quantum theoretical description can lead to a finite
ground state energy at zero temperature. In our classical
approach $T$ plays the role of a parameter in the
momentum distribution employed, to adjust the ground state energy.
}. 
The Momenta are renormalized such that the total initial kinetic
energy of the particles in one nucleus agrees with half of the correct mass 
of 16 nucleons ($M=8\cdot m_n = 8\cdot 939$ MeV) while we assume
that half of the nucleon energy is carried by glue fields. 
This procedure fixes the temperature parameter $T$ in
the Boltzmann distribution for the particles.

\noindent 
The initial configuration of a single nucleus in its rest frame is obtained
through the evolution of the equations
(\ref{Eq.1} -- \ref{Eq.3}), (\ref{Eq.10}), and
(\ref{Eq.11}) over a long period of time starting from the  
initial fields ${\cal E}_{x,k}(t=0) = 0$, ${\cal U}_{x,k}(t=0) = {\bf 1}_2$. 
The simulation of a collision requires a Lorentz boost of each nucleus
into the center of velocity frame of both nuclei. In the example presented
below, the kinetic energy is 100 GeV/u for which $\gamma = 106.5$.
Both nuclei are mapped into their initial positions on a large 
Lorentz-contracted lattice right
after the boost of particle coordinates 
$x_i, p_i$ and field amplitudes.

\noindent 
At this point, it has to be emphasized that it is not possible to boost
fields on a lattice. We have extensively explored the possibility to
generate the color fields before the Lorentz-boost. In this case
one would generate the initial state of the nucleus before the boost
by evolving the equations of motion in time until an equilibrium
between kinetic particle energy and field energy is accomplished.
This would require to employ the algorithm without the
modification mentioned in section 2.
The Lorentz-transformation of such a ``initial state'', however,
does not lead to a configuration which propagates stably along the lattice.
One reason is that the lattice dispersion relation is not 
boost invariant. After the boost, the Fourier components of the fields
do not match with the finite discrete Fourier spectrum of the lattice.
Another reason is that the Hamiltonian of the system does not commute 
with the boost operator.
We have found it most practical to boost the particle coordinates 
$x^{\mu}_i, p^{\mu}_i$ 
at time $t=0$ for both nuclei, wherafter the fields are generated 
from the initial conditions ${\cal E}_{x,k}(t=0) = 0$ and 
${\cal U}_{x,k}(t=0) = {\bf 1}_2$. 
It has been argued \cite{Harris.96} that the coupling constant
$\alpha_s$ should be taken at an effective momentum scale
on the order of $2\pi T$ corresponding to $g=2$.
Starting from $g=0$, the coupling 
constant is increased slowly until it reaches a final value $g=2$.
This method leads to quasi-stable configurations propagating over distances 
which extend over several thousand lattice points in the longitudinal 
direction. It avoids the problem with the lattice dispersion since
the discretized k-space is fixed after the boost. The adiabatic 
increase of the coupling constant avoids the excitation of instable
high frequent modes in the color fields \cite{Gong.94}.    

\noindent 
However, it requires large initial distances between the
nuclei because the charge fluctuation $\mu$ converges slowly
against a final constant value before the collision. Also,
the fields should carry enough energy. 
For nucleons with the correct mass ($m_n = 939$ MeV) 
about half of the total energy are carried by glue fields.
If we want to satisfy this condition to a better approximation
then also semi-hard gluons have to be taken into account. 
The largest fraction of these $50\%$ is carried by soft gluons
with a Bjorken scale $x\le 0.1$, followed by semi-hard gluons
with $0.1<x<0.2$. In the original work
of McLerran and Venugopalan \cite{Venu.94}, only the valence quarks
were taken to be sources of color charge which gave
$\mu^2\sim A^{1/3}\,{\rm fm}^{-2}$. Hence
$\mu \gg \Lambda_{QCD}$ only for nuclei much larger than
physical nuclei. However, if semi-hard gluons 
are also included as 
sources of color charge (as they should be) then $\mu^2$ is 
defined as \cite{Gyulassy.97}
\begin{equation}
\label{Eq.3.2}
\mu^2={{A^{1/3}}\over{\pi r_0^2}}\int_{x_0}^1dx
\Bigl( {1\over{N_c}}q(x,Q^2)+{{N_c}\over{N_c^2-1}}g(x,Q^2)\Bigr)
\end{equation}
where $q,g$ stand for the nucleon quark and gluon structure
functions at the resolution scale of $Q$ of the physical
process of interest. Using the HERA structure function data,
Gyulassy and McLerran estimated that $\mu \le 5\,{\rm fm}^{-1}$
for LHC energies and $\mu \le 2.5\, {\rm fm}^{-1}$ at RHIC
energies. Since semi-hard gluons appear at $x$ values 
which correspond to rapidities greater than the central
rapidity in nuclear collisions, we have to describe also
the longitudinal dynamics of the fields in the collision.
Such a description is possible on a gauge lattice with
a finite extension into the longitudinal direction as
used in the subsequent calculation. For a resolution of the longitudinal
dynamics, the fact that the nuclei move with a speed smaller 
than speed of light becomes very important. This situation
is depicted in the following Fig. 1.
%
%
%
%
%
\begin{figure}[H]
\centerline{
\epsfysize=8cm \epsfxsize=8cm                  
\epsffile{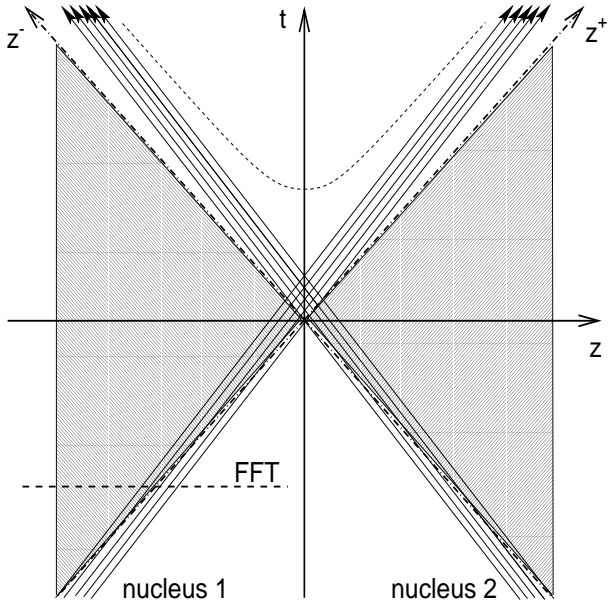}
}
\vskip 0.3cm
\caption{The trajectories of two colliding nuclei (thin solid lines) 
are displayed together with the light cone (dot dashed lines). 
All components of the nuclei stay in the light cone throughout
the collision since they propagate slower than speed of light.
Consequently, the nuclei have a finite extension in the z-direction. 
The dashed horizontal line indicates the direction in which we
carry out the Fourier analysis of the initial state of a single nucleus. }
\end{figure}
\noindent
The finite longitudinal extension of the nuclei is also
necessary for the propagation of the fields on the
lattice.

\noindent 
In the following, we demonstrate that the proposed description
allows to generate an ``initial state'' with a finite
but small charge fluctuation.
For the above chosen lattice constants each nucleon is covered by
40 lattice cells of the dual lattice. Each of these boxes is initially
occupied by an even number $n_b$ of color charged particles
such that the sum over all charges in a box is zero. The initial
charge density thus is zero on a length scale of the lattice spacing
and larger.
At the initial time $t_0$ both nuclei, i.e. the particles, are
boosted with the same $\gamma$ but into opposite directions
towards the center of collision. The lattice constants $a_l$ and
the time step width $\Delta t$ are Lorentz transformed accordingly.
The initial distance $d_0$ of the
nuclei is chosen $16000$ lattice spacings. 
$d_0$ is Lorentz contracted to $45.07$ fm 
in the center of velocity frame. According to the above fixed
extensions into transverse directions and size of the nuclei,
we use a lattice that comprises a total of $4^2\times 16384$ points.
We repeat the calculation for a longitudinally refined lattice which
comprises a total of $4^2\times 32768$ points. 
In this case, the propagation starts at $d_0=32000$ lattice spacings while 
the longitudinal lattice spacing is reduced to $a_3=0.1\,{\rm fm}$,
i.e. a single nucleus is then covered by $4\times 4\times 120$ lattice cells.

While the nuclei translate on the lattice, we adiabatically increase the 
coupling constant with a rate of $\Delta g/\Delta t \le (3000a)^{-1}$ 
until $g$ has reached its final value ($g=2.0$).
The propagating particles transport color charges from one box to
another and the charge density becomes different from zero and
shows fluctuations in coordinate space.
The spacial charge density on the lattice is defined by
\begin{equation}
\label{Eq.3.3}
\rho^c_x(t) := \sum_{i\in I_x(t)} {{q_i^c(t)}\over{a_1a_2a_3}}, 
\end{equation}
where $I_x(t)$ denotes the set of particle indices of the
particles which are in the dual box of the site $x$ at time $t$.
The upper index denotes the color while the lower index refers to
the lattice site at which the density is evaluated.
In Fig. 2, the distribution $d(t,q^c_x)$
of the charge per box is displayed at several
different time steps $t_n$. The charge with color $c$ per box 
at site $x$ is defined as
\begin{equation}
\label{Eq.3.4}
q_x^{c}(t) := \rho^{c}_x(t) a_1a_2a_3
\end{equation}
At time $t=0$ all $q_x^c(t=0)$ are zero and $d(t,q^c_x)$
is different from zero only in the bin around $q_x^c=0$.
In the continuum limit $a_l\rightarrow 0$ this would
correspond to $d(t=0,q_x^c)\sim\delta (q_x^c)$.
From such a configuration we can generate
a configuration possessing a finite charge fluctuation or a 
finite $\mu$ respectively. 
%
%
%
%
%
\vskip 0.2cm
\begin{figure}[H]
\centerline{
\epsfysize=8cm \epsfxsize=8cm                  
\epsffile{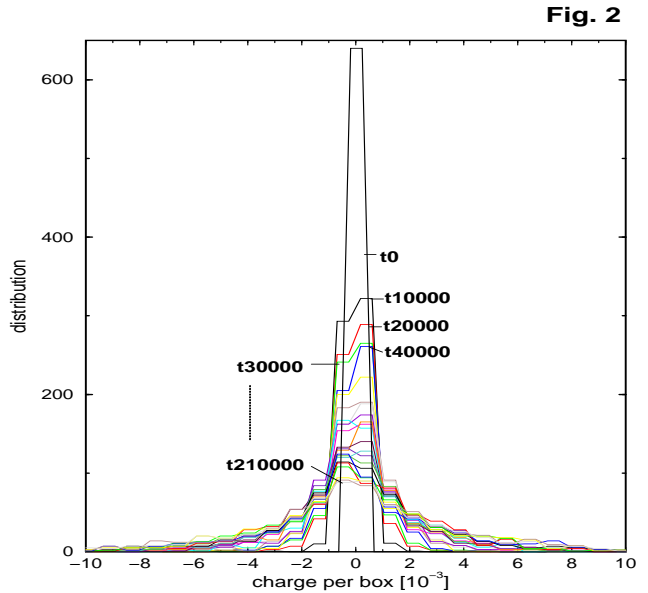} 
}
\vskip 0.3cm
\caption{The computed distributions of
the charge per box 
for a boosted nucleus 
($\gamma = 106.5$, $4\times 4\times 40$ dual lattice cells, 
$n_b=6$, $a_1=a_2=a_3=0.3\,{\rm fm}$)
for increasing times.
The time is indicated at the curves by
the time step numbers $t_n$. 
}
\end{figure}
As already mentioned
above, the initial positions, momenta and color charge
orientations are chosen randomly. The chaotic nature of the
Yang-Mills equations enhances strongly the convergence
towards color distributions in position space which fluctuate
in a highly statistical manner, i.e. the information about the
initial distributions becomes strongly washed out after some
time.

Fig. 2 shows for one color, that the distribution 
of the charge density increases in width monotonously with time.
The corresponding width is displayed in Fig. 5.
Surprisingly, the shape of the distributions 
seems to be different from Gaussian distributions.
This difference is explained in the following way. 
The shape of the distributions in Fig. 2 results from
a superposition of two Gaussian distributions of different
widths which become equal after long times.
In order to show that this effect results from the Lorentz
boost, we display in Fig. 3 the color charge per box distribution
of an unboosted nucleus. These distributions become Gaussian
after relatively short times as indicated by the small time
step numbers $t_n$ in the figure. 
%
%
%
%
%
%
\vskip 0.3cm
\begin{figure}[H]
\centerline{
\epsfysize=8cm \epsfxsize=8cm                  
\epsffile{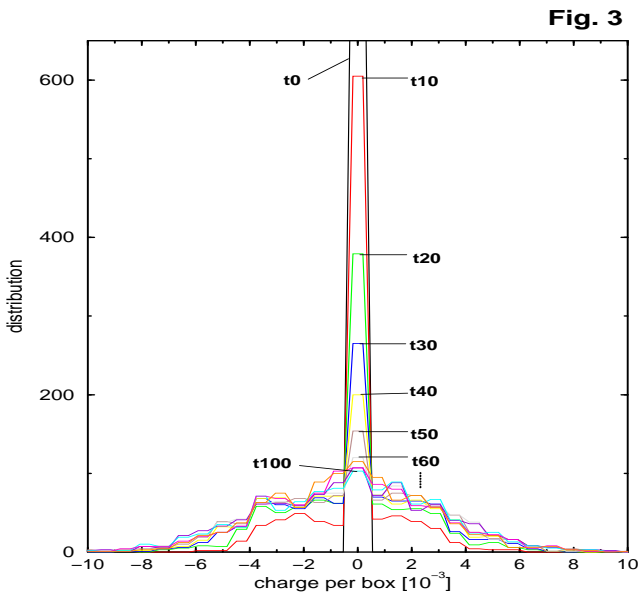} 
}
\vskip 0.3cm
\caption{The same distributions as in Fig. 2 but for an unboosted
nucleus ($10\times 10\times 10$ dual lattice cells, $n_b=6$,
$a_1=a_2=a_3=0.3\,{\rm fm}$).
}
\end{figure}
\noindent
For a boosted nucleus, the lattice is Lorentz contracted in the 
longitudinal direction. The randomly generated coordinates
of the particles imply a color charge fluctuation which appears
after short times in the distributions. In addition to the 
propagation of the charges, a rotation in color space occurs
at a charge when it crosses a cell boundary. 
These rotations lead to an additional smoother broadening of the charge
distributions in each color. The rotation depends
on the length of the link and the strength of the color electric
field pointing in the direction of the boundary transition. 
On a Lorentz contracted lattice, charges that cross transverse
boundaries experience a much stronger rotation than charges 
which cross longitudinal boundaries because the transverse links are 
longer and the transverse color electric fields are stronger
than the longitudinal components. Indeed, for small times we find 
a shape of the charge distribution similar
to that in Fig. 2 in the time evolution of an unboosted collection
of particles on a (by arbitrary choice) deformed lattice.

The curves in Fig. 2 from $t_{10000}$ to $t_{40000}$ contain
a slightly broadened high ``peak'' in the center and a wide
part at the bottom. In order to show that the slow increase of the
width of the central peak results from longitudinal color rotations
we refine the lattice in the longidutinal direction by a factor three
as already mentioned above
while keeping the volume of the nucleus constant. 
The number of particles increases by a factor three while
the length of the charge vector is reduced by a factor three.
%
%
%
%
%
\vskip 0.3cm
\begin{figure}[H]
\centerline{
\epsfysize=8cm \epsfxsize=8cm                  
\epsffile{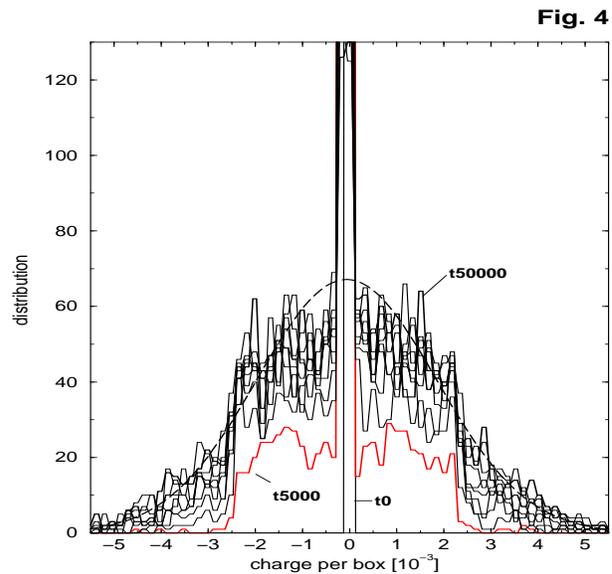} 
}
\vskip 0.3cm
\caption{The same distributions as in Fig. 2 but
for a refined lattice
($\gamma = 106.5$, $4\times 4 \times 120$ dual lattice cells, 
$n_b=6$, $a_1=a_2=0.3\,{\rm fm}$, $a_3=0.1\,{\rm fm}$).
The dashed curve shows the Gaussian fit of the distribution
at time step $t_{50000}$.  
}
\end{figure}
\noindent 
In Fig. 4, we display the color charge distributions for 
a boosted nucleus in the case of a finer longitudinal lattice.
The central peak decays gradually in time but stays narrow
in comparison with the wide distribution at the bottom
and in comparison with the central peak in Fig. 2.
The distribution displayed for the time step $t_{5000}$ results
essentially from charge propagation. The width of this distribution
corresponds exactly to the length $q_0$ of a charge vector. At later
times color rotation seems to smear out the distribution. At 
$t_{50000}$ which is close before the collision, the distribution
has a Gaussian shape. In principle one would have to evolve much
longer in time until the central peak is smeared out. 
The time step numbers in Fig. 4 are much larger (by a factor
$\gamma^2$) than in Fig. 3. This is explained by the fact that
the boosted particles propagate essentially into the longitudinal
direction. In addition, the time step width is reduced by a
factor $\gamma^{-1}$.

In the subsequent
calculations, we neglect the resulting final changes of the width because
they are small.
We find the same time behavior of the distributions for all colors. 
The fluctuation of the charge density $\delta \rho^c_{\rm Latt}$
is associated with the
width of the distributions shown in the figures Fig. 2 and
Fig. 4. Numerically, $\delta \rho_{\rm Latt}$ is determined by
\begin{equation}
\label{Eq.3.5}
(\delta \rho^c_{\rm Latt})^2 := 
{1\over{(a_1a_2a_3)^2}}\sum_x \bigl( \sum_{i\in I_x} q_i^c \bigr)^2,
\end{equation}
where we make use of the fact that
\begin{equation}
\label{Eq.3.6}
\sum_x \sum_{i\in I_x} q_i^c = 0.
\end{equation}
%
%
%
%
%
%
\vskip 0.2cm
\begin{figure}[H]
\centerline{
\epsfysize=8cm \epsfxsize=8cm                  
\epsffile{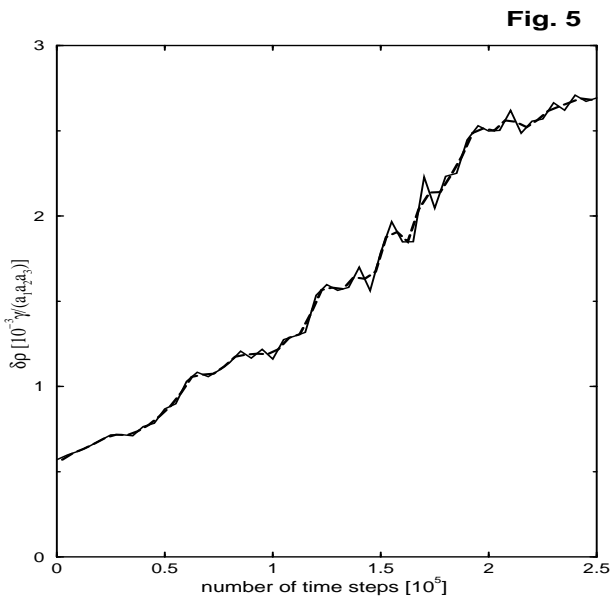} 
}
\vskip 0.3cm
\caption{
The fluctuation of the charge density $\delta\rho_{\rm Latt}$ on the
lattice (upper solid curve) is displayed for a Lorentz contracted
nucleus (see units on axis of ordinates) 
as a function of time over the number of time steps.
The time step width is $\Delta t=0.03/\gamma\,{\rm fm}$.
The dashed curve shows the same but averaged over two points.
}
\end{figure}
In Fig. 5, we display $\delta \rho^c_{\rm Latt}$ of the distributions
shown in Fig. 2 as a function of time. 
Fig. 5 shows, how $\delta\rho^c_{latt}$ evolves after the Lorentz boost
when the nucleus propagates along the contracted lattice.
The curves show that $\delta \rho^c_{\rm Latt}$
begins to converge at large times.
In Fig. 6, we display $\delta \rho^c_{\rm Latt}$ of the distributions
shown in Fig. 4 as a function of time. The curves are plotted
for all colors. 

We now make an estimate of the transverse average charge 
density $\mu$ used in the McLerran Venugopalan model \cite{Venu.94}.
The correct way to determine this quantity is rather involved
since $\mu$ is related to a surface charge density in the
transverse plane. The calculation of this surface charge
density requires a parallel transport on the SU(2) mainfold
of all charges in the longitudinal direction from the 
spacial site of the charge into a fixed transverse plane, e.g.
the central transverse plane of a nucleus. The surface charge
density which follows from that procedure leads to a
distribution of charges per plaquette in the transverse plane.
For reasonable statistics, the transverse extension of the
lattice has to be large enough. 
The transverse extension of our lattice however is small
and we therefore give a simple estimate. We assume that, according
to translational symmetry of the unboosted three dimensional lattice,
the distributions shown in Fig. 2 and Fig. 3 are equal to
the distribution which we would obtain for each transverse layer of
boxes from a lattice with large transverse extension.
The convolution over all transverse charge density distributions
leads to the surface charge density distribution in one 
transverse plane. Supposing that the distributions have a Gaussian
shape, we determine $\mu$ by
\begin{equation}
\label{Eq.3.6b}
\mu = {1\over{N_c^2-1}}\sum_c\delta \rho^c_{latt}a_3\sqrt{a_1a_2{\tilde N_3}}.
\end{equation}
$\tilde N_3$ denotes the number of lattice cells which
cover a nucleus in the longitudinal direction and corresponds to the
number of convolutions.
%
%
%
%
\begin{figure}[H]
\centerline{
\epsfysize=8cm \epsfxsize=8cm                  
\epsffile{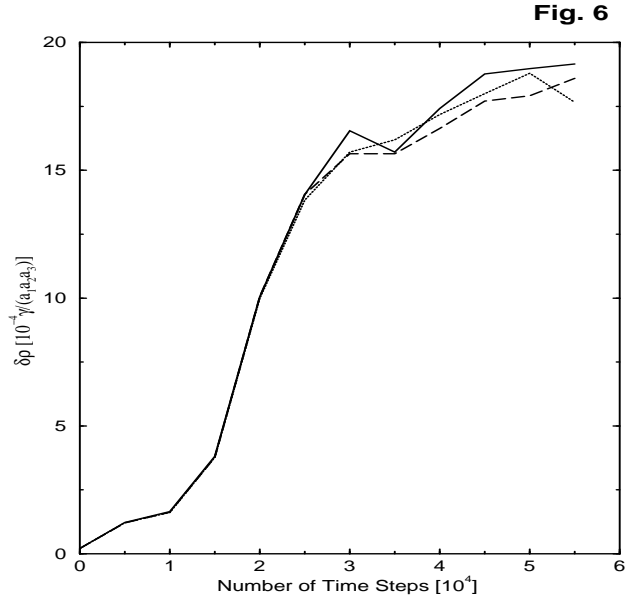} 
}
\vskip 0.3cm
\caption{Same as Fig. 5 but for a lattice with $4\times 4\times 2^{15}$
points and constants $a_1=a_2=0.3\,{\rm fm}$, $a_3=0.1\,{\rm fm}$.
A nucleus is covered by $4\times 4\times 120$ dual lattice cells. 
}
\end{figure}
\noindent
In the calculation of the curves in Fig. 5 and Fig. 6,
we have used $n_b = 6$. According to Fig. 5 and Fig. 6, we obtain
the final values $\mu=0.06\,{\rm fm}^{-1}$ and $\mu=0.04\,{\rm fm}^{-1}$. 
These values stay below the estimate of Gyulassy and McLerran 
\cite{Gyulassy.97}.
The value of $\mu$ or $\delta\rho$ respectively is one important 
criterion for the construction of the
initial state before the collision. Since the charge $Q_0=\sqrt{3/4}$ is
fixed for SU(2) quarks and $g=2$ according to \cite{Harris.96}, 
$n_b$ and the transverse lattice constants $a_1$, $a_2$ are 
the remaining parameters which determine $\mu$.

\noindent 
Another important criterion is that the distribution of 
soft gluons and semi-hard gluons 
should reproduce the longitudinal momentum distribution
$g(x,Q^2)$ which is observed experimentally.
To reproduce $g(x,Q^2)$ would be far too ambitios 
but at least we have to consider the momentum distribution. 
The gluon distributions in momentum space are simply
related to the Fourier transformed gauge field $A_{\nu}^c(\vec k)$
by $\vert 2{\rm Tr}( {\cal A}_{\nu}(\vec k) {\cal A}_{\nu}(\vec k) )\vert$
(no summation over $\nu$.
The light cone model of McLerran and collaborators
describes the initial state through pure gauge fields
\begin{eqnarray}
\label{Eq.3.7a}
A^{\pm} &=& 0,  \\
\label{Eq.3.7b}
A^i &=& \Theta(z^-)\Theta(-z^+)\alpha_1^{i}(x,y)+ \nonumber\\
    & & \Theta(z^+)\Theta(-z^-)\alpha_2^{i}(x,y)\quad i= 1,2.
\end{eqnarray}
\noindent
The coordinates $z^{\pm}$ are defined as 
$z^{\pm} = {1\over{\sqrt{2}}}(t\pm z)$.

\noindent 
In the following, we call the direction $l=3$ which is parallel to
the collision axis the ``longitudinal'' direction and $E,A$-field
amplitudes associated with longitudinal links are called the
``longitudinal fields'' $E_{\rm L}$. 
The fields $E_{\rm T}$ in transverse directions are 
called ``transverse''.
The Fourier transformation of the field of one nucleus
along the dashed line in Fig. 1 leads to a distribution
of the form $\vert A_i(k_3)\vert^2 \sim {1/{k_3^2}}$.
The electric field follows from the equations
(\ref{Eq.3.7a}) and (\ref{Eq.3.7b}) by the
definition $ E=-\partial_t A$. 
The transverse electric field along the dashed line Fig. 1 has the
z-dependence $E_i\sim\delta(z-z_0)\,(i=1,2)$ while
the longitudinal electric field $E_z =0$ or $E^{\pm}=0$
respectively. The Fourier spectrum of the color electric field  
is therefore a constant and shows no $k_3$-dependence.
The following results presented in this section have been calculated
for a single nucleus which propagates on a lattice
of the size $4^2\times 16384$ points.
In Fig. 7, we display as a 
function of the Fourier index $\tilde k_3 = k_3 a_3/(2\pi) $
at the time step $t_{80000}$,
the longitudinal Fourier spectrum of the 
transverse gauge field which we calculate according to 
\begin{equation}
\label{Eq.3.8a}
f_{\rm T} ^{(A)}(t,k_3) :=
{1\over{2 N_1 N_2 (N_c^2-1)}} 
\sum_{x,y}\sum_c\sum_{l=1,2} A^c_l(t,x,y,k_3).
\end{equation}
The corresponding Fourier spectrum of the transverse color electric 
field is calculated as
\begin{equation}
\label{Eq.3.8b}
f_{\rm T}^{(E)}(t,k_3) :=
{1\over{2 N_1 N_2 (N_c^2-1)}} 
\sum_{x,y}\sum_c\sum_{l=1,2} E^c_l(t,x,y,k_3).
\end{equation}
at the same time step.
The gauge fields $A_l^c(t,x)$ have been determined on
the lattice through time integration of the color electric
fields along the whole time evolution. 
The momentum distribution in Fig. 7 agrees
(if one compares the normalized distributions)  
with the result obtained for $f_{\rm T}^{(E)}(t,k_3)$.
This check tells us that
the gauge fields $A_l^c(t,\vec x)$ displayed in Fig. 9 (a) are true
in a sense that they really correspond to the color electric 
fields.
In Fig. 8, the Fourier spectrum of the
longitudinal gauge field is displayed on a logarithmic scale,
i.e. $\log_{10}(\vert f_{\rm L}^{(A)}(t,k_3)\vert^2k_3^2)$ is plotted
as a function of the Fourier index $\tilde k_3 = k_3 a_3/(2\pi) $.  
The figure shows that the longitudinal components of the fields
carry very low longitudinal momenta in comparison to the longitudinal
momenta in the transverse components. Most of the longitudinal
field energy is in modes with $\tilde k < 34$ while 
the transverse field energy is in modes up to $\tilde k=3400$.

\noindent 
Fig. 9 (a) shows for all colors the gauge fields generated
by a nucleus which moves from the left to the right 
in the figure. The steep front of the field distribution
indicates the position of the nucleus.
%
%
%
%
%
\begin{figure}[H]
\centerline{
\epsfysize=8cm \epsfxsize=8cm                  
\epsffile{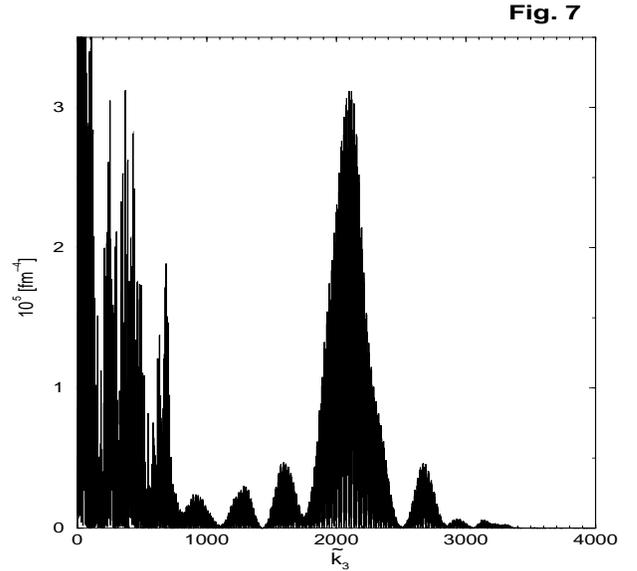} 
}
\vskip 0.3cm
\caption{
Longitudinal Fourier-spectrum of the transverse 
gauge fields $\vert f_{\rm T}^{(A)}(t_{80000},k_3)\vert^2 k_3^2$ 
in units of $10^{5}\,{\rm fm}^{-4}$ plotted as a function of
the Fourier index $\tilde k_3$.
}
\end{figure}
\noindent 
Lets assume that the color electric field $E_l^c(z)$
is finite and constant inside of the nucleus and zero outside of
the nucleus, i.e. $E_l^c(z)=1$ for ${{-\Delta z}\over{2}} \le
z \le {{\Delta z}\over{2}}$ and $E_l^c(z)=0$ for 
$z > {{\Delta z}\over{2}}$ or $z < {{-\Delta z}\over{2}}$.
The Fourier transformation of such a static rectangular field
distribution has the form
\begin{equation}
\label{3.9}
\tilde E_l^c(k_3) = {{2^{3/2}}\over{\pi}}
{{\sin{( {{\Delta z}\over2}k_3 )}}\over{k_3}}
\end{equation}
from which we conclude that the half period of
$\vert E_l^c(k_3)\vert^2$ in $k$-space is 
$\Delta k_3 = {{2\pi}\over{\Delta z}}$. This difference
can also be expressed in terms of the Fourier index
as $\Delta\tilde k = {{N_3a_3}\over{\Delta z}}$.
For a moving nucleus, the Fourier spectrum is shifted as
can be seen in Fig. 7. The width of the large hump in
Fig. 7 corresponds to the longitudinal extension of the 
boosted nucleus which is $\Delta z = 40 a_3 $ while
$N_3=16384$.
%
%
%
%
%
\begin{figure}[H]
\centerline{
\epsfysize=8cm \epsfxsize=8cm                  
\epsffile{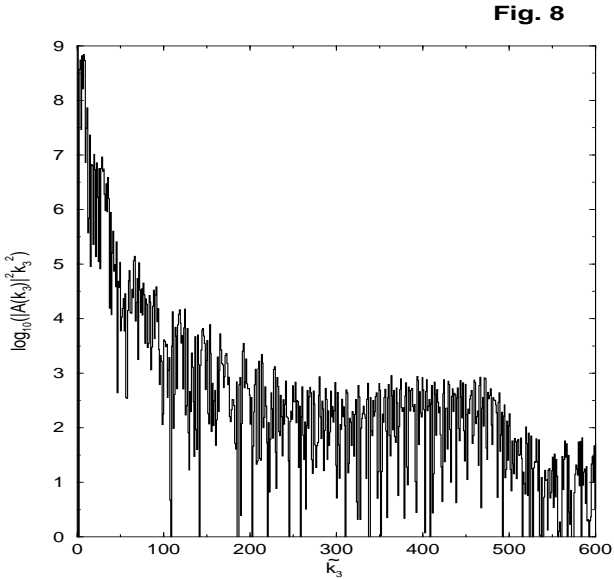} 
}
\vskip 0.3cm
\caption{The longitudinal Fourier-spectrum of the longitudinal 
gauge fields is shown as a function of the Fourier index 
$\tilde k_3$ on the logarithimic scale as  
$2\log_{10}{\bigl( \vert f_{\rm L}^{(A)}(t_80000,k_3)\vert k_3\bigr)}$.
The argument of the logarithm is given in units of ${\rm fm}^{-2}$.
}
\end{figure}
\noindent 
In Fig. 9 (a), we display a snap shot of the gauge field amplitudes
$\vert\overline A^c(t,z)\vert^2$ taken as average over the transverse planes
at the time step $t_{100000}$.
Fig. 9 (b) shows for the same time step
the corresponding transverse energy density
distribution which we define as  
\begin{equation}
\label{Eq.3.10a}
w_{\rm T}^{(E)}(z) = 
Tr({\cal E}_1 {\cal E}_1) + Tr({\cal E}_2 {\cal E}_2).
\end{equation}
Further below, we will also need the longitudinal energy density
distribution which is defined
\begin{equation}
\label{Eq.3.10b}
w_{\rm L}^{(E)}(z) = Tr({\cal E}_3 {\cal E}_3).
\end{equation}
The figures Fig. 9 (a) and Fig. 9 (b) clearly show that the gauge
field behind the moving color sources 
is mostly a pure gauge field because the amplitude
of the color electric field behind the nucleus is zero.
Large amplitudes of the color electric field appear only
inside of a nucleus which has a finite extension in the longitudinal
direction. The gauge fields displayed in Fig. 9 (a) correspond to the
gauge fields of the light cone model according to the
ansatz in Eq. (\ref{Eq.3.7a}).
In the light cone model, behind a nucleus,
they are constant along lines parallel to the collision axis.
These fields, however, are not unique due to gauge freedom.
Therefore, the gauge fields in Fig. 9 (a) can be very different from the
ansatz in Eq. (\ref{Eq.3.7a}).
%
%
%
%
\vspace{0.5cm}
\begin{figure}[H]
\centerline{
\epsfysize=8cm \epsfxsize=8cm                  
\epsffile{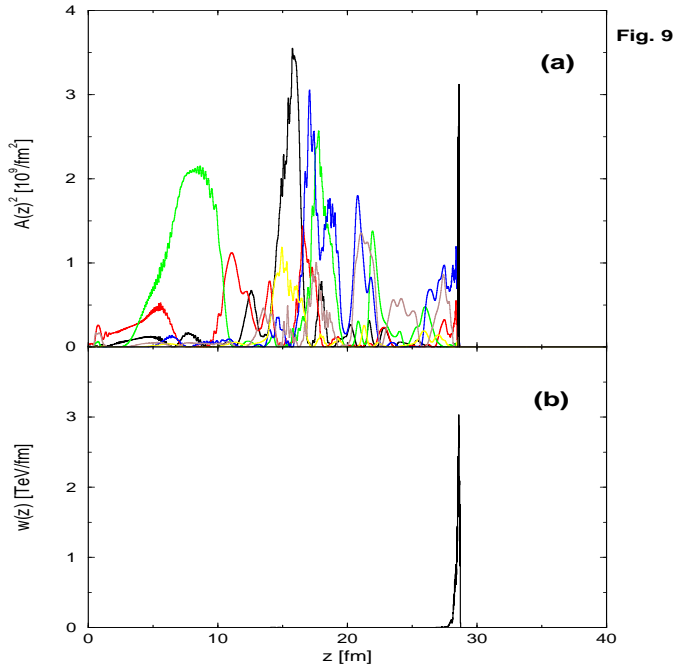} 
}
\vskip 0.5cm
\caption{
The transverse gauge fields $\vert\overline A^c(t_{100000},z)\vert^2$ 
averaged over the transverse plane and squared 
are displayed for each color in the upper panel.
In the lower panel, we show the corresponding transverse color
electric field energy density $w_{\rm T}^{(E)}(t_{100000},z)$.  
}
\end{figure}
%
%
%
%
%
%
%
%
\section {The Collision}
%
%
\noindent 
In the previous section, we have outlined how a collision can
be simulated on a gauge lattice with small transverse extension
and how the initial state can be generated within the model.
In this section, we focus on the time period after the collision
and discuss results
obtained from simulations of a collision on two different
lattices. In one case we performed a calculation on a lattice
of the size $4\times 4\times 2^{15}$ points using lattice
constants $a_1=a_2=0.3\,{\rm fm}$ and $a_3=0.1\,{\rm fm}$.
Accordingly, a single nucleus is covered by $4\times 4\times 120$
dual lattice cells. In the following the refer to this
paramerization as the ``case (1)''. 
In the second case, we performed a calculation on a lattice
of the size $4\times 4\times 2^{14}$ points using lattice
constants $a_l=0.3\,{\rm fm}\,(l=1,2,3)$. 
A single nucleus is thus covered by $4\times 4\times 40$
lattice cells. We refer to this parametrization as the
``case (2)''.
A time-step width of $\Delta t = 0.03/\gamma$\,fm is used
in both cases.  
In the case (2), we construct initial states as
``realistic'' as possible within the model. We try to ajust
the glue field energy of the nucleon, to obtain a color charge
fluctuation $\mu$ in the range estimated by other authors
and to describe the longitudinal size of the Lorentz boosted
nucleus. We expect that this calculation provides a 
non-perturbative estimate for soft glue field scattering
and glue field radiation at high energies. 
Due to the high color field energies the simulation
of case (2) is close to the limit of acceptable numerical precision.
An improvement
requires finer but also larger lattices exceeding our
computational resources. Instead, we refine the lattice
in case (1) keeping the size small. In this case the fields
carry less energy before the collision but the resolution
of the field dynamics is higher and more precise. Case (1)
therefore provides a control over the qualitative nature of the     
results while we expect some quantitative results from case (2). 
Most of the figures presented below are shown for case (1). The
curves agree in a qualitative manner with results obtained in case (2) 
but they show less intensity.

\noindent 
In Fig. 9 (b) of the previous section, we have shown the 
transverse color electric
field energy density distribution $w_{\rm T}^{(E)}(z)$ 
of a propagating nucleus for one time step.
To get a full picture of the evolution of $w_{\rm T}^{(E)}(t,z)$ and
$w_{\rm L}^{(E)}(t,z)$ during the generation of the initial state
when the two nuclei approach each other, we display in Fig. 10 (a) and
(b) the energy densities of the transverse and longitudinal  
components of the $E$-field every 10000 time steps for the case (1). 
\vspace{0.5cm}
%
%
%
%
\begin{figure}[H]
\centerline{
\epsfysize=8cm \epsfxsize=8cm                  
\epsffile{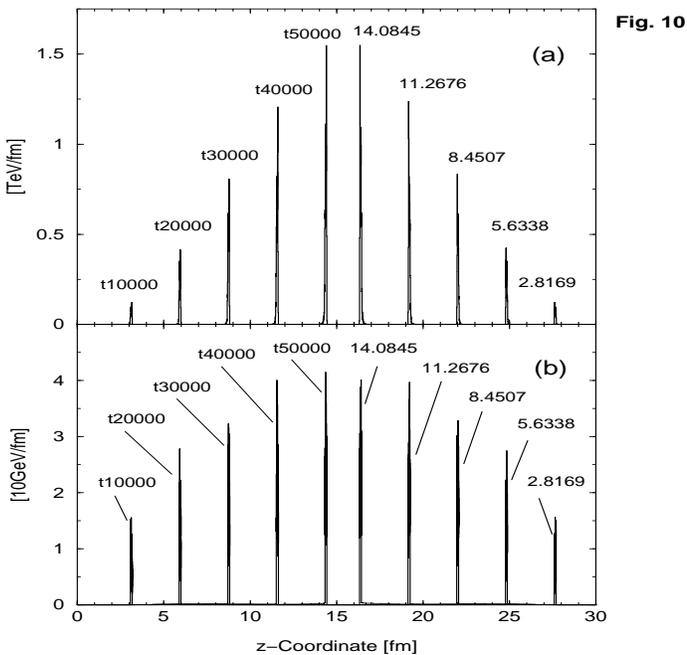} 
}
\vspace{0.5cm}
\caption{The transverse (a) and longitudinal (b) energy densities
of the color electric fields are displayed for two approaching
nuclei. The numbers on top of the distributions indicate the
number of steps in the time integration (on top of left nucleus)
or time in units of fm (on top of right nucleus).   
}
\end{figure}

\noindent 
We verified, that the corresponding $B$-field energy densities
$w^{(B)}(z)$ are almost identical with $w^{(E)}_{\rm T}(z)$. After
1000 time steps (not shown in the figure)
the transverse $E$-field energy $W^{(E)}_{\rm T}$ is by about a 
factor $\gamma/2$ larger than the longitudinal $E$-field 
energy $W^{(E)}_{\rm L}$. This ratio increases to $\gamma$ at 50000 
time steps and remains constant afterwards. 

\noindent 
As Fig. 10 (a),(b) shows, two configurations representing two nuclei 
propagate remarkably stably (nucleus 1 from left to right and
nucleus 2 from right to left) over long distances on the lattice. 
The collision begins at $t_{53333}$ where the fields of both
nuclei make first contact.

\noindent 
In Fig. 11 at time step $t_{50000}$ of case (1), 
the particle density (in units of 20 on the ordinate), 
which is defined as the number of particles per bin of width $a/10$
in longitudinal direction, is superimposed on the transverse  
and longitudinal field energy densities (multiplied by a factor 10
in the figure) for comparison. 
The particle density defines the extension of the nucleus in
the longitudinal direction. The mismatch between particle density 
and field energy density decreases when the density 
of lattice points covering the nucleus in the longitudinal direction is
increased. This requires a shift of the cutoff (here at $k_c=2.0$ GeV 
for $a=0.3\,{\rm fm}$) to higher momenta.  
%
%
%
%
\begin{figure}[H]
\centerline{
\epsfysize=8cm \epsfxsize=8cm                  
\epsffile{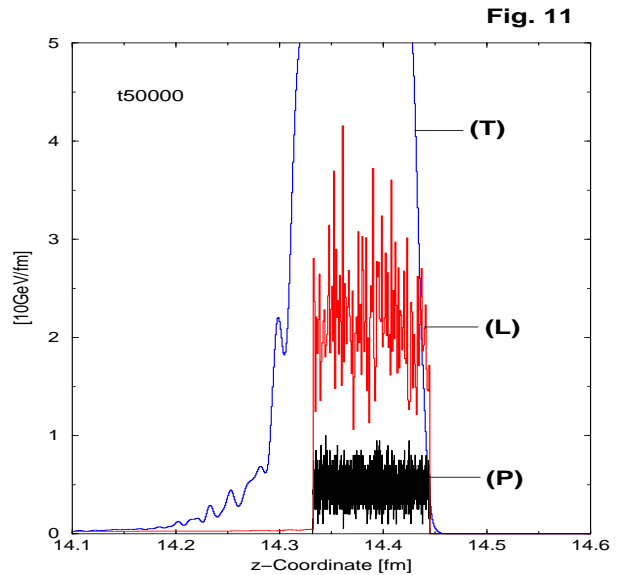} 
}
\vspace{0.3cm}
\caption{Comparison of the transverse (T) and longitudinal (L)
color electric field energy distribution in a single
nucleus with the distribution of the particles. The 
longitudinal energy density has been multiplied by
a factor 10 for better comparison. The particle density (P)
has been divided by a factor 20.
}
\end{figure}
The tails of the field energy densities 
result not only from lattice dispersion effects. 
The lattice is Lorentz contracted in the longitudinal direction and 
particles have large longitudinal momentum components. Therefore, 
particles primarily transfer energy into longitudinal links.
The nonlinearity of the Yang Mills
equations provides a mechanism transferring energy from 
longitudinal into transverse degrees of freedom. Field amplitudes
which are left behind on the longitudinal links are canceled by 
following particles of rotated color charge. 
\vspace{0.5cm}
%
%
%
%
%
\begin{figure}[H]
\centerline{
\epsfysize=8cm \epsfxsize=8cm                  
\epsffile{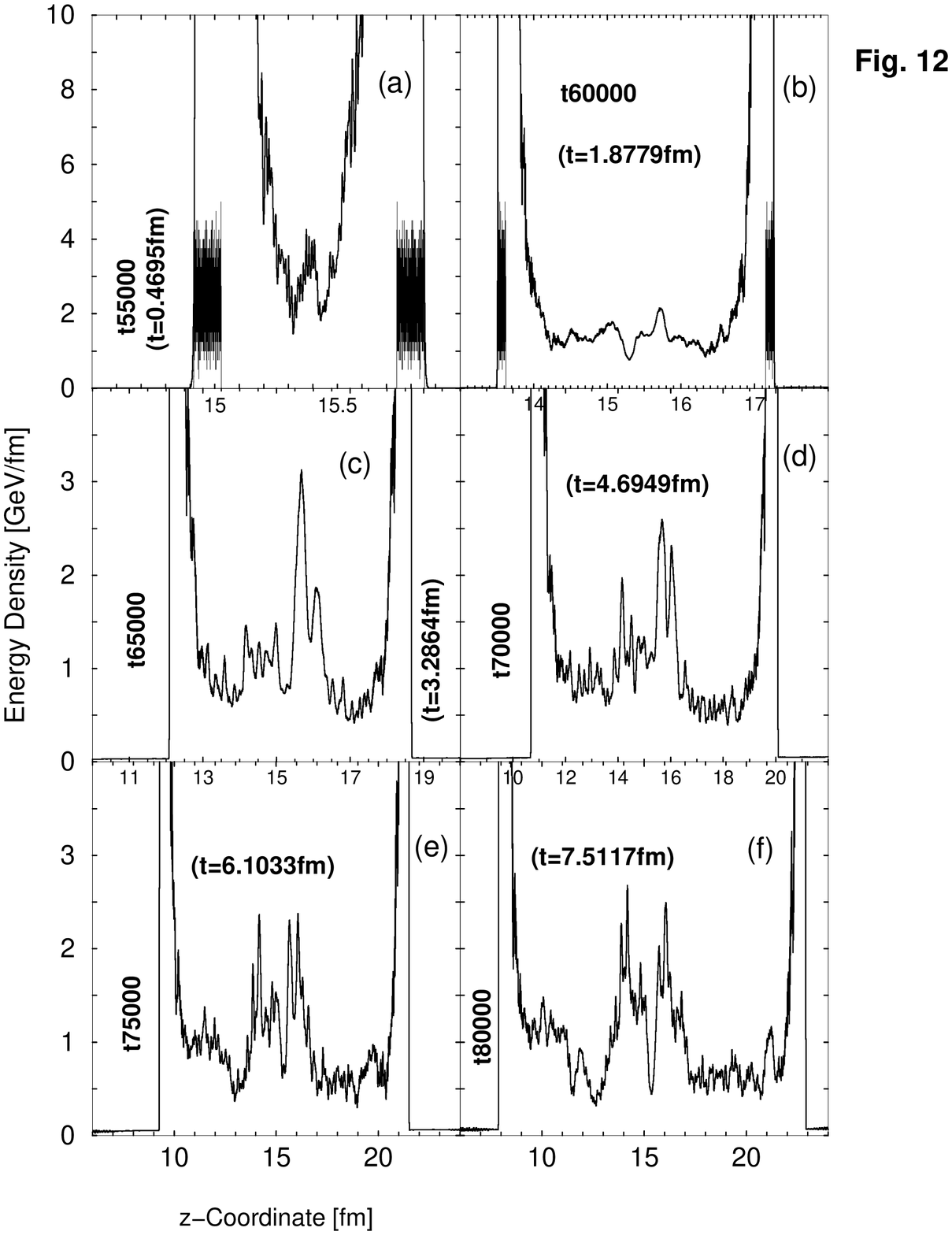} 
}
\vspace{0.5cm}
\caption{The transverse color electric field energy density is
displayed for case (1) at various different time steps after the begin
of the collision. The total time step number $t_n$ and the
relative time $t$ counted from the first contact of the nuclei 
is indicated in each panel.
The collision begins at the total time step $t_{53333}$
or at the relative time $t=0\,{\rm fm}$ respectively. 
The particle distributions (multiplied by a factor 1/10) are included in 
panel (a) and (b). 
}
\end{figure}
\noindent 
Fig. 12 displays snap shots of $w^{(E)}_{\rm T}(t_n,z)$ 
1667, 6667, 11667, 16667, 21667 and 26667 time steps after the 
begin of the collision. 
In the panel (a) and (b) the particle distributions in coordinate space
(multiplied by a factor 1/10) 
are displayed together with the field energy distributions to
indicate the position of the receding nuclei. 

\noindent 
As compared to Fig. 11, the width of the distributions $w^{(E)}_{\rm T}(t,z)$ 
moving with the particles is considerably increased (about $30\%$). 
This sudden
change of the width occurs during the overlap of the nuclei. At the same
time the field amplitudes at the position of the particles have decreased.
We would expect that this leads to an increased energy transfer from the
particles into the fields.
To explore this behavior, we define the electric field energy for each color
\begin{equation}
\label{Eq.4.1}
W_c^{(E)}(t) = {{a_1 a_2 a_3}\over 2} 
               \sum_{x\in X}\sum^3_{l=1}  E^c_{x,l} E^c_{x,l}.
\end{equation}
Note, that we do not sum over the color index c.
The magnetic field energy $W_{\rm T}^{(B)}(t)$
is defined in the same manner. In the calculations, we find
that $W_{c}^{(B)}(t) \simeq W_{c}^{(E)}(t)$.
We calculate the integrated field energy $W_{c}^{(E)}(t)$
as a function of time. 
Fig. 12 will be further discussed below.
%
%
%
%
%
%
\vspace{0.5cm}
\begin{figure}[H]
\centerline{
\epsfysize=8cm \epsfxsize=8cm                  
\epsffile{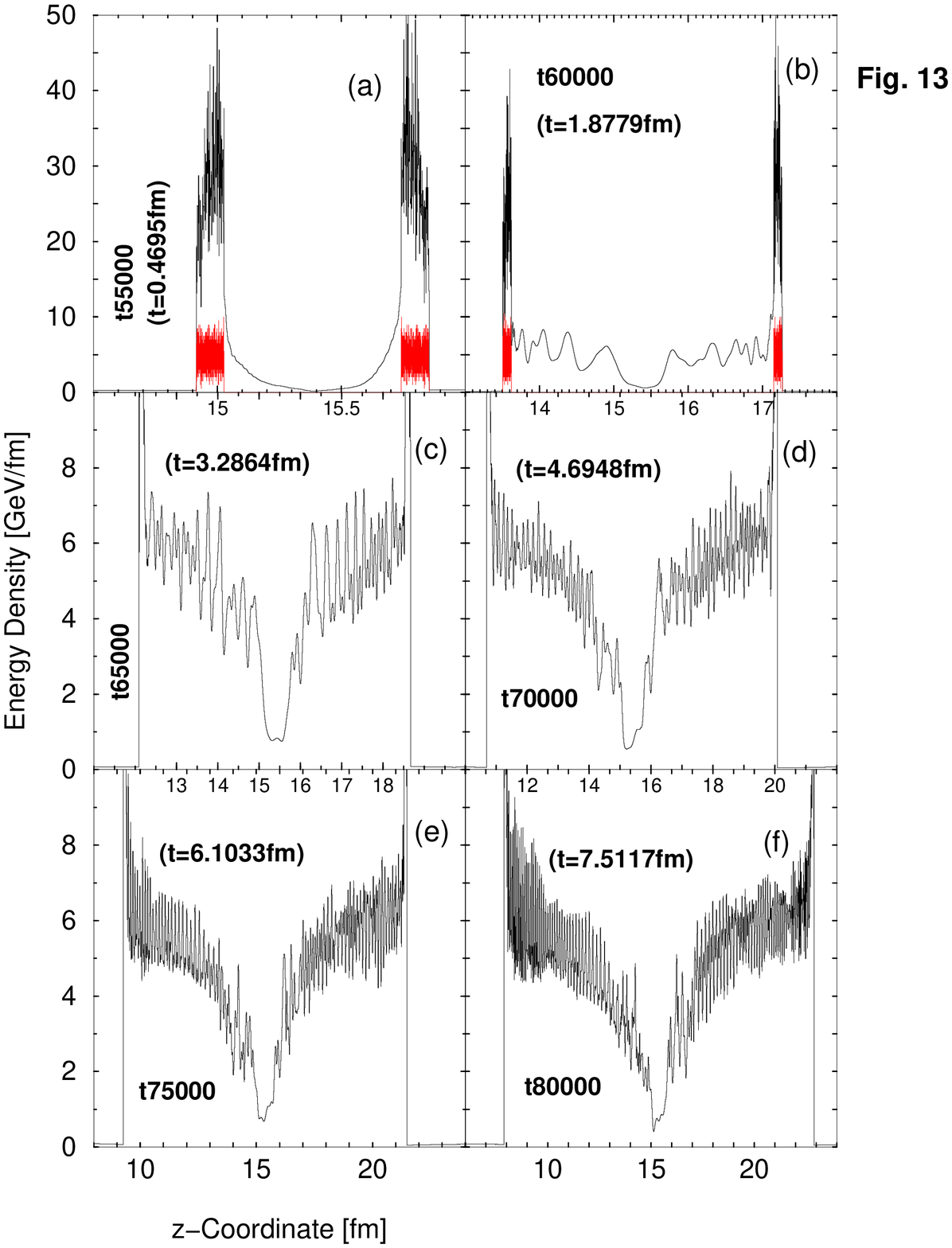} 
}
\vspace{0.5cm}
\caption{Same as in Fig. 12 but for the longitudinal 
color electric field energy density
$w^{(E)}_{\rm L}(z)$.
}
\end{figure}
In Fig. 14, we display for case (2) 
$W_{c}^{(E)}(t)$ for each $c$ in a time interval
around the time of first contact $t_{80000}$.
In the time region between $t_{80000}$ and $t_{90000}$,
we observe a kink (it is a bit washed out) in $W^{(E)}_{c}(t)$, 
which increases remarkably in the collision (see also Fig. 15). 
In the upper left corner, we display for case (1)
$W^{(E)}(t) = \sum_c W_{c}^{(E)}(t)$ in a time interval 
around $t_{53333}$. The angle between the dashed line and the
dot-dashed line indicates the kink. 
The kink in Fig. 14 corresponds to an increased energy transfer
from the particles into the fields while the nuclei recede.
This tells us that the energy deposit between the receding
nuclei results not only from the interaction between the
fields during the overlap time but glue field radiation contributes.
The contributions to glue field radiation appear in the
longitudinal field energy density distributions shown in Fig. 13.
Energy is transfered from the particles into
the fields through longitudinal links. Before the collision
during the generation of the initial state
this transfer occurs smoothly generating coherent fields.
The strong increase of $w^{(E)}_{\rm L}(t,z)$, at times
larger than the time of the kink and further away from the
center of collision, results from the fact that more
energy is transfered into longitudinal links than the closest
transverse links can absorb from them
during the short time in which
a nucleus passes a certain longitudinal link. 
The distribution $w_{\rm L}^{(E)}(z)$ 
increases sharply for collision times larger $0.3\,{\rm fm}$
and arrives at a first maximum at $t\simeq 0.5\,{\rm fm}$.
The next minima appear at a time $t\simeq 0.7\,{\rm fm}$
which could presumably be interpreted as a thermalization time for
the soft glue fields. It corresponds to the FWHM of the valley
in Fig. 13 (c).
%
%
%
%
%
\begin{figure}[H]
\centerline{
\epsfysize=8cm \epsfxsize=8cm                  
\epsffile{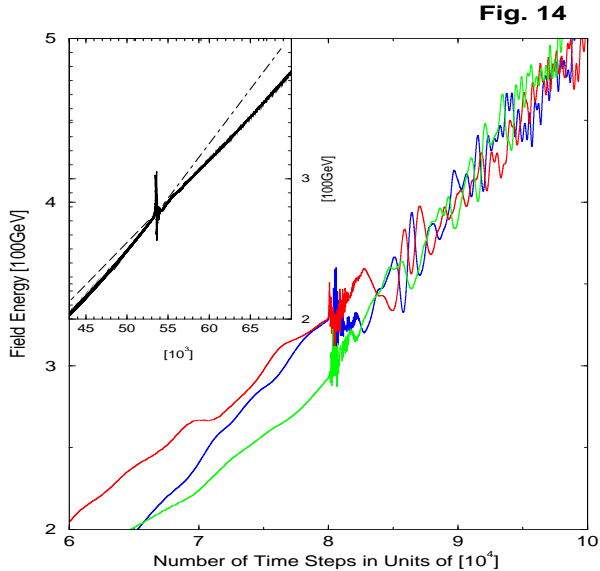} 
}
\vskip 0.3cm
\caption{The total color electric field energy
$W_{c}^{(E)}(t)$ for each color is displayed for case (2) in the
time interval between $t_{60000}$ and $t_{100000}$. The panel
in the upper left corner displays the total electric field enery
for case (1). The lines indicate a kink.
}
\end{figure}
\noindent 
When the two nuclei pass through each other, the particles experience 
the combined fields of both nuclei. The corresponding changes of the
field amplitudes enter into the r.h.s. of (\ref{Eq.3})
and modify the orientation of the color charge vectors $\vec q_i^a(t)$
during the rather short overlap time of $0.11\,{\rm fm}$.
This induces net color charge currents 
resulting in radiation of gluons. Since the longitudinal momenta of
the particles are large compared with their transverse components,
color charges move essentially parallel to longitudinal links
and induce electric fields on these links. Before the collision,
the charge of the following particles was polarized such
that these fields were canceled. After the collision 
this is no longer true over a long period of time. 
A continuation of
the time evolution has shown that the longitudinal energy
density of the fields remains as large as in Fig. 13 and
decreases slowly at very large times.

\noindent 
The excess of energy remaining in the longitudinal links propagates
then into transverse directions. 
These longitudinal fields possess only transverse momenta
and result in gluon radiation into transverse directions. 
This mechanism describes  
gluon radiation from the particles. 
The fact that the
transverse extension of our lattice is very small and
periodically closed does not really allow an expansion
of the radiated glue fields into transverse directions.
They collide with other radiated gluons and are stopped.
This feature is somewhat unphysical and can only be removed
by taking lattices with large transverse extensions.
The radiated glue fields remain therefore essentially
on the longitudinal links and decay slowly. However, at
least it allows one to determine how much is radiated from
a certain volume.

\noindent 
We therefore define the color separated longitudinal
electric field energy
\begin{equation}
\label{Eq.4.2}
W_{{\rm L},c}^{(E)}(t)
                       = {{a_1a_2a_3}\over 2}
                       \sum_{x\in X} E^c_{x,3} E^c_{x,3},
\end{equation}
which is a function of time. Fig. 15 displays
$W_{{\rm L},c}^{(E)}(t)$ for all colors. The kink
in the time dependent field energy of Fig. 14 
appears much stronger in Fig. 15. The curves in
Fig. 15 show essentially the cumulative time integral 
of how much energy 
is radiated from the particles at increasing time.
\vspace{0.5cm}
%
%
%
%
%
\begin{figure}[H]
\centerline{
\epsfysize=8cm \epsfxsize=8cm                  
\epsffile{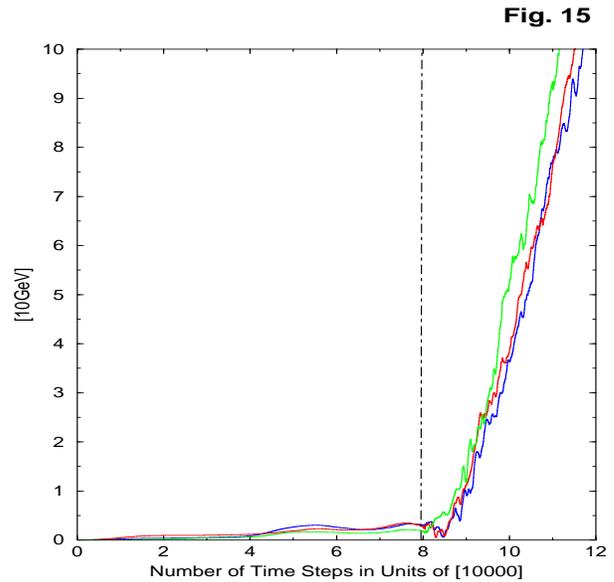} 
}
\vskip 0.3cm
\caption{Longitudinal color electric field energy as
function of time.
}
\end{figure}
\noindent 
In the following, we discuss the contribution to the energy
deposit which results from the interaction of the fields.
The contribution of the particles will be discussed later. 
We discuss the results for two choices of the lattice
size and lattice spacing. The numbers, obtained in
the case where we use a $4\times 4 \times 16768$ lattice
are given in parentheses. 
As we see from Fig. 12, a considerable fraction of the 
transverse field energy is 
deposited around the center of collision between the 
receding nuclei. 

At the time step $t_{160000}$ in the case (2),
about $440\,{\rm GeV}$ 
of the total transverse electric field energy    
$W^{(E)}_{\rm T}= 2810 \,{\rm GeV}$ 
are left in the spatial region 
between $z_1=3.08\,{\rm fm}$ and $z_2=43.08\,{\rm fm}$
(the center of collision is at $z=23.08\,{\rm fm}$). 
The field energy
density between the nuclei is about a factor $1/100$ smaller
compared to the field energy density at the position of the
particles in the receding nuclei. Due to the large space volume 
between $z_1$ and $z_2$ however, the energy deposit amounts
to about $16\%$ of the total energy in soft field modes. 

\noindent 
Fig. 13 shows that the longitudinal energy densities are
small but finite around the center of collision.
The calculation shows in case (2) that within
short times $\Delta t = 1.0\,{\rm fm}$
after the begin of the collision 
about $23\,\%$ ($7.42\,{\rm GeV}$) of the total field
energy deposit ($31.98\,{\rm GeV}$) that is left
in the region between $z_1=22.08\,{\rm fm}$ and $z_2=24.08\,{\rm fm}$
are carried by longitudinal components of the $E$-field. The average ratio 
$w^{(E)}_{\rm L}/w^{(E)}_{\rm T}$ in this region is about 0.6 which is
large compared to the above mentioned factor $1/\gamma$. 
This ratio appears to be even larger in case (1).
Within the short time $\Delta t = 0.3\,{\rm fm}$ after the begin
of the collision about $29\%$ ($664\,{\rm MeV}$)
of the total field energy deposit ($2270\,{\rm MeV}$) left in
the region between $z_1=15.08\,{\rm fm}$ and $z_2=15.68\,{\rm fm}$
are contained in longitudinal component of the $E$-field.

Since the color charge and color current density is zero, the linearized
Yang-Mills equations (\ref{Eq.5}) are homogeneous in this region and
allow only for solutions with non-zero amplitudes into
transverse directions in relation to the direction of their
energy flow. 
Figure 10 displays the fraction
with momenta pointing into transverse direction relative to
the collison axis. 
The longitudinal field energy density in the narrow space region
between $z_1=15.08\,{\rm fm}$ and $z_2=15.68\,{\rm fm}$ is not
a result of gluon radiation.
A calculation with the right-hand side of 
(\ref{Eq.2}) and (\ref{Eq.3})
set to zero just before the nuclei make first
contact yields the same results in that narrow region,
which finds its explanation in a transfer of energy from the propagating 
transverse fields to the longitudinal fields due to the non-linear 
coupling between $E_{\rm T}$ and $E_{\rm L}$ in (\ref{Eq.5}). 
Also in collisions of Yang-Mills wave packets, where no particles
are taken into account, we observed a strong increase of the 
longitudinal fields around the center of collision during the
overlap of the wave packets \cite{Po.98}.

\noindent 
When the two nuclei overlap during the collision, the fields are 
superposed. As a result of the nonlinear terms in (\ref{Eq.5}),
which act as source terms for the longitudinal fields, the
amplitudes $E^c_{\rm L}$ start to grow rapidly during
the short overlap time (of the particle clouds) of $0.11\,{\rm fm}$.
This mechanism describes the scattering of soft gluons which
occurs over a much longer period of time.
To analyze this process within the model, 
we define the following quantities.
The Poynting vector in the adjoint representation is 
denoted as
\begin{equation}
\label{Eq.4.3}
\vec {\cal S} := c\, \vec {\cal E}\times {\cal B}.
\end{equation}
With the transverse and longitudinal components of the vector 
(\ref{Eq.4.3}) we define the total transverse and longitudinal
energy currents
\begin{eqnarray}
\label{Eq.4.4a}
i_T(t):= 2\sum_{l=1}^2 \int d^3x 
          \,\big\vert{\rm Tr}
          \bigl( {\cal S}_l(t,\vec x) \bigr)
          \big\vert
\\
\label{Eq.4.4b}
i_L(t):= 2\sum_{l=3}^3 \int dx^3 
          \,\big\vert{\rm Tr}
          \bigl( 
          {\cal S}_l(t,\vec x)
          \bigr)
          \big\vert .
\end{eqnarray}
In the following Fig. 16, we display $i_T(t)$ (panel (a))
and $i_L(t)$ (panel (b)) for the time interval between
$t_{50000}$ and $t_{70000}$. Fig. 16 (a) shows that
shortly after the first contact of the nuclei there
is a sudden increase of the transverse energy current
in the fields. This increase describes soft gluon
scattering into transverse directions. The slope of the dot-dashed
line indicates the current corresponding to the 
energy rate transfered from the particles
to the fields if no collision would occur.
The dot-dashed line rpresents an underground
which has to be subtracted from the curve to determine the
contribution that results from scattering.
A very similar behavior has recently been found in collisions of
color polarized wave packets on the three-dimensional lattice
\cite{Po.98}.
At the same time, Fig. 16 (b) shows that  
the longitudinal energy current decreases remarkably
during the overlap of the fields. 
The width of the inverse peak in Fig. 16 (b) corresponds
to a time of $0.18\,{\rm fm}$ which is the overlap time of the fields.
After the overlap, the curve approaches the dot-dashed line
almost as close as before the collision. This leads to the
conclusion that the inverse peak is a interference phenomenon.
Nevertheless, it determines the time interval in which the fields
interact. 
The steady increase of the longitudinal current before the
collision results
from the energy transfer from particles into fields.
In a similar study for case (2), we find that
the transverse energy current keeps increasing much stronger
long after the collision. These additional contributions
describe glue field radiation. 
\vskip 0.3cm
%
%
%
%
%
\begin{figure}[H]\centerline{
\epsfysize=8cm \epsfxsize=8cm                  
\epsffile{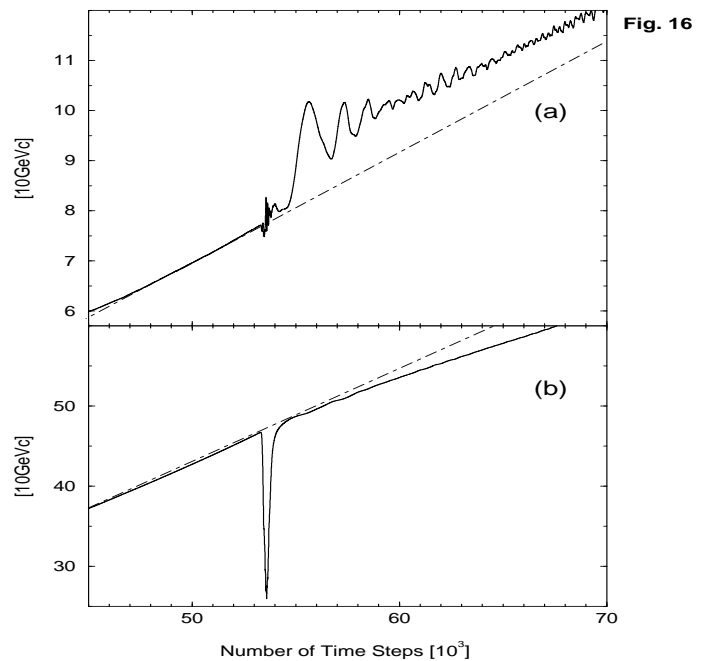} 
}
\vspace{1.0cm}
\caption{The transverse (panel (a)) and 
longitudinal (panel (b)) energy currents
$i_T(t)$ and $i_L(t)$ are displayed in a time interval
around the beginning of the collision.
}
\end{figure}
\noindent 
To give an estimate for glue field radiation, we integrate
the curve in Fig. 13 (f) from $z_1=8.38\,{\rm fm}$ to $z_2=22.38\,{\rm fm}$
and find an energy deposit of $64.62\,{\rm GeV}$. This energy divided
by the volume covered by both particle distributions and divided by the time
of $t=7.51\,{\rm fm}$ since first contact yields a radiation power
of about $115\,{\rm GeV/fm^4}$. In the case (2) we find the result
$110\,{\rm GeV/fm^4}$ over a time of $23\,{\rm fm}$.
We now give estimates for contributions from soft scattering and radiation.
The total field energy at the time step 
$t_{80000}$ amounts to $757.24\,{\rm GeV}$ in case (1) while
the field energy deposit in the space region between
$z_1=8.38\,{\rm fm}$ and $z_2=22.38\,{\rm fm}$
is $94.74\,{\rm fm}$. We conclude that $8\%$ of the total field
energy are deposited between the nuclei at $t=7.51\,{\rm fm}$.
In case (2), we find $26\%$ at $t=23\,{\rm fm}$.

\noindent 
Finally, we analyze the momentum distribution in
the transverse color electric fields. In the previous
section we have discussed the Fourier spectrum of
the initial state of a single nucleus. Now, we transform
the fields of two nuclei propagating at the same time
on a lattice of the size $4\times 4\times 2^{15}$
according to case (1).
In the upper panel (a) of Fig. 17, we display the 
normalized longitudinal momentum distribution of two nuclei
at time step $t_{50000}$ shortly before the beginning 
of the collision.
With Eq. (\ref{Eq.3.8b}) it is defined as
\begin{equation}
\label{Eq.4.5}
\overline f_{\rm T}^{(E)}(t,k_3):=
f_{\rm T}^{(E)}(t,k_3)
\Bigl[ \int_0^{\infty} dk_3 f_{\rm T}^{(E)}(t,k_3) \Bigr]^{-1}
\end{equation}
In the lower panel (b), we display the momentum 
distribution after the collision at the time step $t_{80000}$.
A comparison of Fig. 17 (a) with Fig. 17 (b) shows evidence
for the production of a final state with a momentum
distribution that dramatically differs from that of the
initial state.
\vspace{1.0cm}
%
%
%
%
%
\begin{figure}[H]
\centerline{
\epsfysize=8cm \epsfxsize=8cm                  
\epsffile{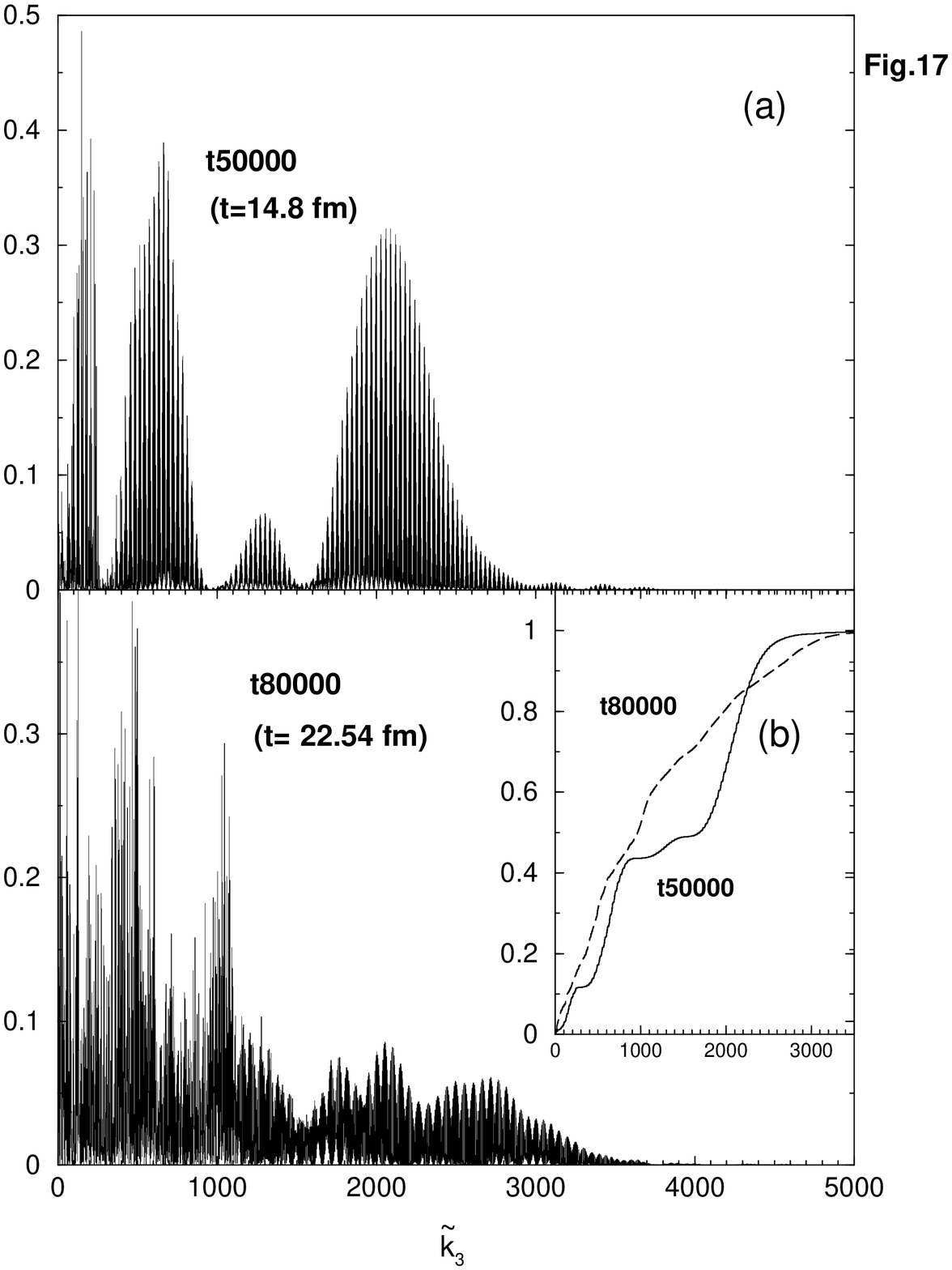} 
}
\vspace{1.0cm}
\caption{The normalizd Fourier spectra of the transverse color electric field
before (a) and after the collision (b). The snap shots are taken
at the time steps $t_n$ indicated on top of the curves. In the upper
left corner of panel (b)), we display the cumulative integrals over the
spectra.
}
\end{figure}
\noindent 
In the upper right corner of Fig. 17 (b), we display the
cumulative integrals over the distributions 
\begin{equation}
\label{Eq.4.6}
\overline F_{\rm T}^{(E)}(t,k_3):=
\Bigl[ \int_0^{\infty} d k_3 f_{\rm T}^{(E)}(t,k_3) \Bigr]^{-1}.
\int_0^{k_3} dk'_3 f_{\rm T}^{(E)}(t,k'_3)
\end{equation}
We compare
the results for $t_{50000}$ (solid curve) 
and for $t_{80000}$ (dashed curve).
Above $\tilde k = 2200$, the dashed curve is clearly below
the solid curve which shows that energy
is transfered from high frequent modes into low
frequent modes during the collision. This behavior has
also been found in collisions of color polarized 
Yang-Mills wave packets in 1+1 dimensions \cite{Hu.95}
and later in collisions of Yang-Mills wave packets 
in 3+1 dimensions \cite{Po.98} and could presumably
describe particle production \cite{Gong.94}. Our calculation
shows that this pure classical phenomenon occurs also
in collisions of Yang-Mills fields which are not polarized
in color space. 
%
%
%
%
%
%
\section {Summary and Conclusion}
\noindent 
\noindent 
In the framework of a classical effective approach of QCD,
color charged particles have been employed to generate the
color source effects of the valence quarks in ultra-relativistic
nuclear collisions.
Classical Yang-Mills fields have been used to describe soft modes 
of the gluons.  
We have studied the time-evolution of such a system 
of particles and fields
by solving the coupled system of
Wong equations and Yang-Mills equations, 
as a transport model for the
reaction of nuclei in collisions at high energies. 
We have carried out numerical simulations of the collision
on a long three dimensional gauge lattice with small
transverse extensions. 
Results have been presented which demonstrate the properties of
such a description from various aspects and provide estimates
for contributions at small transverse momenta. 
We have discussed the similarities and the 
basic differences of the approach
presented in this paper as compared to the light cone source model
\cite{Venu.98,Ayala.95,Jalilian.1,Jalilian.2,Jalilian.3,Rischke.97}.
It has been demonstrated that a ``initial state''
can be generated which describes the longitudinal
extension, the color charge fluctuation and the glue
field energy of the nuclei before the collision.
The shape of the longitudinal momentum distribution 
of the soft gluons in the initial state as obtained
in our calculation is similar to
the shape of the experimentally know gluon distribution at small $x$.

\noindent 
As a function of time, we have calculated the
transverse and longitudinal energy densities 
of the color electric fields.
For an effective coupling of $g=2$ and
for an energy of $100\,{\rm GeV/u}$, 
our calculation predicts that a
relatively large fraction (over $8\%$) of the total soft  
field energy (at small $x$) contributes at $t=7.5\,{\rm fm}$
after the first contact to the energy deposit
between $z_0\pm 7\,{\rm fm}$ around the center of collision $z_0$.

\noindent 
The colliding color fields cause soft gluon scattering.
This has been demonstrated in a
calculation of the transverse and the longitudinal energy current
as a function of time throughout the collision.

\noindent 
The non-linear nature of the Yang-Mills equations causes
a change of the color charge distribution in
the two colliding clouds of particles during the overlap
time which results in an increase of the charge fluctuation. 
The resulting color charge currents lead to an averaged gluon radiation
from the particles which amounts to $110\, {\rm GeV/fm^4}$
(radiation power per particle volume) 
over times of $7\,{\rm fm}$ and even $20\,{\rm fm}$.

\noindent 
The longitudinal momentum distributions of the 
transverse color electric fields have been calculated.
The fields were not polarized in color space.
Our results show that an energy transfer from
high frequenzy modes into low frequenzy modes occurs
in the collision. The same observation has been made 
by Hu et al. \cite{Hu.95} studying collisions of color polarized 
pure Yang-Mills wave packets on a one-dimensional SU(2) gauge 
lattice. The Fourier analysis of section 5 shows that the existence of 
this phenomenon does not require color polarization in the
initial conditions, e.g. before the collision. 

\noindent 
In this paper we have discussed a new approach to describe 
ultra-relativistic heavy-ion collisions in a coherent way
in 3+1 dimensions. We have presented results from numerical
model simulations to explore the possibilities for
describing results from experiments. A combination of our
approach with parton cascades and ultra-relativistic 
quantum molecular dynamics could allow to trace further
back from experimental data into the early stage of the
collision. The approach for itself could describe the
formation of the quark gluon plasma within the classical
mean field approximation.

\noindent 
At the present stage, we are still far from accomplishing
such goals. It would be important to increase the extension
of the lattice into transverse directions to cover two
nuclei completely. This would allow to determine the soft
transverse momentum distributions and the time evolution
of single transverse modes. 
Also a study of the dynamics of the particles will be
necessary. This requires that the particles are fully
coupled to the fields. The momentum distributions and
rapidity distributions of the particles could be compared
with experiment. \newpage
\noindent 
We thank S.G. Matinyan, D.H. Rischke, and R. Venugopalan for 
helpful discussions.
This work was supported by the U.S. Department of Energy under
Grant No. DE-FG02-96ER40495.



\end{document}